\documentclass[11pt]{revtex4-2}
\usepackage[T1]{fontenc}
\usepackage[utf8]{inputenc}
\usepackage{graphicx}
\usepackage{dsfont}
\usepackage{xcolor}
\usepackage{amsmath}
\usepackage{amssymb}
\usepackage{mathtools}
\usepackage{amsfonts}
\usepackage{bbold}
\usepackage{braket}

\begin{document}
\preprint{}
\title{Absorption cross section in gravity's rainbow from confluent Heun equation}

\author{Kihong Kwon}
  \email{rlghd0926@dgu.ac.kr}
  \affiliation{Division of Physics and Semiconductor Science, Dongguk University,
    Seoul 04620, Republic of Korea} 
\author{Juli\'{a}n Barrag\'{a}n Amado}
  \email{julian.barragan.amado@gmail.com}
  \affiliation{Division of Physics and Semiconductor Science, Dongguk University,
    Seoul 04620, Republic of Korea}
\author{Bogeun Gwak}
  \email{rasenis@dgu.ac.kr}
  \affiliation{Division of Physics and Semiconductor Science, Dongguk University,
    Seoul 04620, Republic of Korea}
  
\begin{abstract}
We investigate the scattering of a massless scalar field by a charged non-rotating black hole in the presence of gravity's rainbow.
Using the connection coefficients of the confluent Heun equation expressed in terms of the semi-classical confluent conformal blocks and the instanton part of the Nekrasov-Shatashvili (NS) free energy, we obtain an asymptotic expansion for the low-energy absorption cross section. 
\end{abstract}



\maketitle
	
\section{Introduction}

General Relativity (GR) is the theory of space, time, and gravitation formulated by Einstein in 1915. However, general relativity is a purely classical theory, and one would expect that if the principles of quantum theory are to apply to the gravitational field, general relativity must be an effective theory at low energies of a truly fundamental theory of gravity, where the scale at which the classical description breaks down is determined by the Planck length  $l_{P}$ \cite{Wald:1984rg}. There are several candidates for a quantum theory of gravity including string theory \cite{zwiebach2004first, polchinski1994string}, loop quantum gravity \cite{rovelli2008loop, nicolai2005loop} and non-commutative geometry \cite{aschieri2006noncommutative}. Nevertheless, these theories should reproduce GR predictions at low energies.

Doubly (or deformed) special relativity is a class of modified special relativity theories that expect that global Lorentz symmetry emerges as an approximate symmetry at low energies from the quantum theory of gravity \cite{rainbow1}. 
It suggests that the relativistic energy-momentum relation is changed to a nonlinear form
\begin{equation}
    c^4m^2=E^2-c^2p^2+f(E,p^2; E_{P}),
\end{equation}
where the energy-momentum relation depends on the energy of the particle \cite{amelino2002doubly,amelino2002relativity}. Furthermore, the Lorentz transformations are modified to a nonlinear form as well \cite{magueijo2003generalized, magueijo2002lorentz}. Such modified dispersion relation leads to the gravity's rainbow. This theory is not a UV completion of general relativity because this approach assumes $E \ll E_{P}$, where $E$ is the energy of the quanta and $E_{P}$ is the Planck energy. Moreover, it should recover classical general relativity for $E/E_{P}\rightarrow0$. According to this theory, the space-time will depend on the energy of the quanta.

Black holes are very compact objects, predicted by Einstein's general relativity, and have an event horizon and a singularity. They are unique laboratories to test general relativity in the strong-field regime, where one can expect quantum effects to become relevant. Furthermore, we can also set bounds on the validity of alternative theories of gravity using experimental results from gravitational wave detection \cite{will2004testing, harko2009testing, cardoso2017testing, zhou2018testing}.
By the same token, the scattering of linear field perturbations by a black hole can provide insights on the astrophysical phenomena; namely, the bending of light rays and the (in)elastic cross sections. In particular, one can see that in the low-energy regime the absorption cross section of a black hole is given by the area of the horizon \cite{gibbons1975vacuum,das1997universality,crispino2009scattering}.

Recently, the problem of linear perturbations of black hole solutions has attracted renewed interest regarding the appearance of linear second-order ordinary differential equations after the separation of variables. Typically, the resulting ODEs possess a finite number of regular (irregular) singular points depending on the background and the presence of horizons, e.g., the dynamics of massless scalar fields in Kerr-(anti)-de Sitter black holes will reduce to two Heun equations \cite{Novaes:2014lha}, where the connection between the Heun accessory parameter and the four-point Virasoro conformal blocks \cite{Litvinov:2013sxa} brought a new application for the computation of quasi-normal modes (QNMs) \cite{CarneirodaCunha:2015hzd,CarneirodaCunha:2015qln,Novaes:2018fry,BarraganAmado:2018zpa} in terms of an isomonodromic tau function formulated in \cite{Gamayun:2012ma}. The latter step was made under the Alday-Gaiotto-Tachikawa (AGT) correspondence, where the semi-classical limit of the Liouville CFT corresponds to the Nekrasov-Shatashvili limit of a $4D$ $\mathcal{N}=2$ $SU(2)$ supersymmetric gauge theory \cite{Alday:2009aq}. In particular, the four-point conformal block of Liouville primary fields on the Riemann sphere is described by the Nekrasov partition functions of $SU(2)$ gauge theory with $N_{f}=4$ \cite{Aminov:2020yma}. Furthermore, the confluent limits of the Heun equation have been identified with correlation functions of vertex and irregular vertex operators, providing a new way to study exact solutions of second-order ODEs of the Fuchsian type using the appropriate two-dimensional CFT. In this context, the long-standing connection problem for (confluent) Heun equations has been solved in terms of a connection problem for semi-classical conformal blocks in \cite{Bonelli:2021uvf,Bonelli:2022ten}. These ideas have been applied to study QNMs, tidal Love numbers, and greybody factors \cite{Consoli:2022eey,Bianchi:2021mft,Bianchi:2021xpr,Bianchi:2023sfs} and thermal two-point functions in holography \cite{Dodelson:2022yvn}.

The present work aims to investigate the low-energy absorption cross section of a Reissner-Nordstr\"{o}m black hole in gravity's rainbow by exploiting the connection coefficients of the Confluent Heun Equation. Although some effort has been devoted to studying the emission and absorption properties by rotating and charged black holes \cite{maldacena1997universal,jung2004proof,crispino2009scattering,leite2017absorption,Gwak:2021tcla,Gwak:2022nsia}, and the scattering coefficients have been thoroughly investigated in alternative theories of gravity \cite{gup1,Anacleto:2019tdj,anacleto2020absorption,Gwak:2022mzea,anacleto2023absorption}, our results show the rainbow effect on the asymptotic expansion for the $s-$wave absorption cross section in the small $\omega$ limit \eqref{eq:seccion}. As we will see, higher order corrections will be related with the instanton part of the NS free energy of a $SU(2)$
$\mathcal{N}=2$  gauge theory with $N_{f}=3$ fundamental hypermultiplets.


The paper is organized as follows. In Section \ref{sec:2}, we will review gravity's rainbow and introduce the background. Section \ref{sec:3} comprises the bulk of the results. First, we analyze the Klein-Gordon equation for a massless scalar field, and then, we separate the radial ODE into the near and far from the black hole, which will allow us to solve it analytically in each region. Next we match the solutions in the overlaping region to find the conserved fluxes across the horizon and incoming from infinity. We compare the contribution of arbitrary rainbow functions to the absorption cross section in Section \ref{sec:3.2}. In Section \ref{sec:3.3} we provide the explicit computation of the absorption cross section in terms of the connection coefficients of the Confluent Heun equation. Finally, we discuss our results in Section \ref{sec:4}. The definition of the NS free energy can be found in Appendix \ref{appendix_A}.  

\section{Review on gravity's rainbow}
\label{sec:2}
\noindent
In the theory of Special Relativity, the dispersion relation is given by
\begin{equation}\label{eq:dispersion}
    E^2-\left(pc\right)^2=\left(mc^2\right)^2,
\end{equation}
and from now on, we set $c=\hbar=G=1$, where $G$ is the universal gravitational constant. To describe the physics at extremely high energies, one can find several candidates. Among them, the deformed or doubly special relativity suggests that the dispersion relation is changed to an energy dependent modified dispersion relation \cite{rainbow1},
\begin{equation}
    E^{2}f(E/E_{P})^{2} - p \cdot p \,g(E/E_{P})^{2} = m^{2},
\end{equation}
where $E$ is the energy of a quanta measured by the observer, and $E_{P}$ means the Planck energy, which is constant for all inertial observers. This can be regarded as acting as a nonlinear map from the momentum four-vector to itself,
\begin{equation}
    U\cdot (E, p_i)=(U_0, U_i)=\left (f\left(\frac{E}{E_{P}}\right)E,\ g\left({\frac{E}{E_{P}}}\right)p_i\right ),
\end{equation}
where the functions $f(E)$ and $g(E)$ are energy dependent functions, and represent the dependence of the modified dispersion relation with respect to the energy of the quanta. These functions are sometimes called rainbow functions \cite{ali2015proposal, darlla2023black}. In the limit $E/E_{P}\rightarrow 0$, $f(E)$ and $g(E)$ converge to $1$ to recover the ordinary dispersion relation \eqref{eq:dispersion}. 
There are several rainbow functions proposed in the literature, and as examples, we single the following \cite{awad2013nonsingular,amelino1997potential,magueijo2002lorentz}:
\begin{equation}\label{ex1}
    f(E)=\frac{e^{E/E_{P}}-1}{E/E_{P}}, \quad g(E)=1,
\end{equation}
and
\begin{equation}\label{ex2}
    f(E)=g(E)=\frac{1}{1-E/E_{P}},
\end{equation}
considered by Amelino-Camelia and Magueijo and Smolin, respectively. We will consider generic rainbow functions to see the effect of gravity's rainbow on the absorption cross-section.

We assume that free wave solutions in flat space-time are preserved as plane wave forms \cite{rainbow1, rainbow2}. That is, 
\begin{equation}
    dx^ap_a=dx^0p_0+dx^ip_i.
\end{equation}
Thus, the dual position space is transformed to the following form \cite{rainbow1,rainbow2}:
\begin{equation}\label{eq:example2}
    d\tilde{x}^0(E)\rightarrow\frac{1}{f(E/E_{P})}dx^0, \quad d\tilde{x}^i(E)\rightarrow\frac{1}{g(E/E_{P})}dx^i,
\end{equation}
where $d\tilde{x}^{\mu}$ and $dx^{\mu}$ denote the basis one-form of the energy dependent and independent frame, respectively. 
In the energy-independent frame, one can see the dependence of metrics with respect to the rainbow functions. Thus, if an observer sees particles in the same place but having different energies, the observer concludes that each particle is assigned a different metric, or if two other observes see a particle but each one measures different energies, they will assign different metrics depending on the value they have measured. Such a one-parameter energy-dependent set of metrics is called the `rainbow metric'.

Now, let us consider a spherical and static space-time. Following the same procedure in \cite{rainbow1}, the spherical symmetric and static solution has the following form: 
\begin{equation}\label{eq:rainbow_metric}
ds^2=-\lambda(r)\frac{dt^2}{f(E)^{2}}+\frac{1}{\lambda(r)}\frac{dr^2}{g(E)^{2}}+\frac{r^2}{g(E)^{2}}d\Omega^2, \quad (d\Omega^2=d\theta^2+\sin^2{\theta}d\phi^2).
\end{equation}
Here, the radial coordinate $r$ is defined as the area coordinate, which is proportional to the square root of a physical area observed at particular radius. Thus, we see the radial coordinate measured by a quanta of energy $E$ is $\tilde{r}=r/g(E)$. On the other hand, the angular coordinates $\theta$ and $\phi$ are always energy independent, so only the time and radial coordinates can have energy dependence. 

In this paper, we consider the Reissner-Nordstr\"{o}m (RN) black hole with gravity's rainbow. Therefore, the function $\lambda(r)$ and the electric potential in the energy independent frame are expressed as \cite{Gim:2018axz}
\begin{equation}\label{eq:roots}
\lambda(r) = 1-\frac{2M}{r}+\frac{g(E)Q^2}{r^2}, \qquad A_{\mu}dx^{\mu} = -\frac{g(E)}{f(E)}\frac{Q}{r}dt.
\end{equation}
The location of the inner and outer horizon is given by the roots of $\lambda(r) = 0$
\begin{equation}\label{eq:horizons}
r_\pm=M\pm\sqrt{M^2-g(E)Q^2}.
\end{equation}
Note that in the energy independent coordinate, the area of the horizon has an additional rainbow function $g(E)$ in the denominator, hence if we measure it at $r=r'$ using some quanta with energy $E$, we will find an energy dependent area, $A=4\pi r'^2/g(E)^{2}$. As a result, we can think of the radius given by $r=r^{\prime}/g(E)$, which is consistent with the idea in \eqref{eq:example2}.

\section{Absorption cross section}
\label{sec:3}

\subsection{Scalar wave equation}
\label{sec:3.1}

The Klein-Gordon equation for a massless scalar field in the metric \eqref{eq:rainbow_metric} is given by
\begin{equation}\label{eq:Klein_Gordon}
\frac{1}{\sqrt{-g}}D_{\mu}\left(\sqrt{-g}g^{\mu\nu}D_{\nu}\right)\Psi = 0,
\end{equation}
with $D_{\mu} = \nabla_{\mu} - i e A_{\mu}$, $g$ is the determinant of the metric, and $e$ is the charge of the field. Then, the equation \eqref{eq:Klein_Gordon} can be explicitly written as follows
\begin{align}\label{eq:KG}
    &\frac{1}{\sqrt{-g}} \partial_\mu(\sqrt{-g}g^{\mu\nu}\partial_\nu\Psi) \nonumber\\
    &=-\frac{f(E)^{2}}{\lambda(r)}\left(\partial_t - i e A_{t}\right)^{2}\Psi+\frac{2g(E)^{2}}{r}\lambda(r)\partial_r\Psi+g(E)^{2}\frac{d\lambda(r)}{dr}\partial_r\Psi\nonumber\\
        &+g(E)^{2}\lambda(r)\partial_r^2\Psi +\cot{\theta}\frac{g(E)^{2}}{r^2}\partial_\theta\Psi+\frac{g(E)^{2}}{r^2}\partial_\theta^2\Psi+\frac{g(E)^{2}}{r^2\sin^2{\theta}}\partial_\phi^2\Psi=0.
\end{align}
Since the metric \eqref{eq:rainbow_metric} is static and spherical symmetric, we can introduce the following ansatz \cite{berti2009quasinormal}:
\begin{equation}\label{eq:ansatz}
\Psi_{\omega, \ell, m}(t,r,\vartheta,\varphi) = e^{-i\omega t}Y_{\ell}^{m}(\vartheta,\varphi)R_{\omega,\ell}(r),
\end{equation}
where $\omega$ is the angular frequency of the field without the rainbow function because the plane wave solution is preserved \cite{rainbow1, rainbow2}. It turns out that the latter separates \eqref{eq:KG} into two second-order ODEs: The angular part can be written in terms of the scalar spherical harmonics $Y_{\ell}^{m}(\theta, \phi)$ on the 2-sphere, with the eigenvalues determined by the equation
\begin{equation}\label{eq:angular_ODE}
\left[\frac{1}{\sin\theta}\partial_\theta\left(\sin\theta\partial_\theta\right)+\frac{1}{\sin^2\theta}\partial_\phi^2\right]Y_\ell^m(\theta, \phi) = -\ell(\ell+1)Y_\ell^m(\theta, \phi),
\end{equation}
where $\ell$ and $m$ are integers satisfying $\ell \ge 0$ and $|m|\le\ell$. The radial part is given by
\begin{equation}\label{eq:radial_ODE}
\left[\frac{d}{d r}\left(r^{2}\lambda(r)\frac{d}{d r}\right) + \frac{r^{2}}{\lambda(r)}\left(\frac{f(E)}{g(E)}\omega - \frac{e Q}{r}\right)^{2} - \ell(\ell+1)\right]R_{\omega,\ell}(r) = 0,
\end{equation}
and for convenience we define $\hat{\omega} = \frac{f(E)}{g(E)}\omega$.

We will proceed as follows: in section \ref{sec:3.2}, we decompose the problem in two regions, near and far from the black hole. The latter assumption allows us to simplify the radial ODE and write an analytical solution in both cases, while section \ref{sec:3.3} is devoted to addressing the full solution of the radial ODE using the results of the Confluent Heun equations and their connection coefficients presented in \cite{Bonelli:2022ten}. Nevertheless, the computation of the absorption cross section is given by the ratio between the going flux into the horizon and the flux coming in from infinity \cite{maldacena1997universal}. Hence, we need to impose the following boundary conditions on the radial solutions around the singular points involved in the scattering:

\begin{equation}\label{eq:boundaryforr}
R_{\omega,\ell}(r) \sim
\begin{cases}
A_{in}e^{-i\hat{\omega} r}r^{-1-i\left(\hat{\omega}(r_{+}+r_{-})-e\,Q\right)} + A_{out}e^{i\hat{\omega} r}r^{-1+i\left(\hat{\omega}(r_{+}+r_{-}) - e\,Q\right)}, & r \rightarrow \infty\, \\
\\
\hfil A_{tr}(r-r_{+})^{-i\frac{r_{+}^{2}}{r_{+}-r_{-}}\left(\hat{\omega} -\frac{e Q}{r_{+}}\right)}, & r \rightarrow r_{+}. 
\end{cases}
\end{equation}

\subsection{The overlapping method}
\label{sec:3.2}

We will consider the wave solution in the low-energy limit, i.e., $\omega\ll 1/M$, which means the Compton wavelength of the particle is much larger than the size of the black hole. Furthermore, we define the near-region $r - r_{+} \ll 1/\omega$ and the far-region $M \ll r-r_{+}$. 

\subsubsection{The near-horizon equation and solution}
\label{sec:3.2.1}

In the near region, the radial equation \eqref{eq:radial_ODE} for $e = 0$ reduces to
\begin{equation}\label{eq:radial_near}
r^2\lambda(r)\frac{d}{d r}\left(r^2\lambda(r)\frac{d}{d r}R(r)\right) + r_+^{4}\hat{\omega}^{2}R(r) - r^2\lambda(r)\ell(\ell +1)R(r) = 0.
\end{equation}
In order to solve the near-region equation, we introduce a new variable
\begin{equation}
z = \frac{r-r_+}{r-r_-},\qquad 0 \le z \le 1.
\end{equation}
As a result, we have
\begin{equation}
    (r-r_+)(r-r_-)\partial_r=(r_+-r_-)z\partial_z
\end{equation}
and, the equation \eqref{eq:radial_near} becomes
\begin{equation}\label{eq:near_equation}
z(1-z)\partial_z^2R(z)+(1-z)\partial_zR(z)+\frac{\hat{\omega}^2 r_+^4}{(r_+-r_-)^2}\frac{1-z}{z}R(z)-\frac{\ell(\ell + 1)}{1-z}R(z)=0
\end{equation}
where
\begin{equation}
    P = \frac{r_+^2}{r_+-r_-}.
\end{equation}
We can define the following s-homotopic transformation
\begin{equation}
    R(z) = z^{i\hat{\omega}P}(1-z)^{\ell+1}\phi(z),
\end{equation}
which brings the equation \eqref{eq:near_equation} into the form 
\begin{equation}\label{eq:rad_hypergeometric}
z(1-z)\partial_z^2\phi(z)+[1+2i\hat{\omega}P-(1+2i\hat{\omega}P+2(\ell+1))z]\partial_z\phi(z)-((\ell+1)^2+2i\hat{\omega}P(\ell+1))\phi(z)=0,
\end{equation}
where the solutions are given in terms of the hypergeometric functions \cite{arfken1999mathematical}, and we can identify the parameters
\begin{equation}
    a = 2i\hat{\omega}P+\ell+1,\qquad b = \ell+1,\qquad c = 1+2i\hat{\omega}P.
\end{equation}
The two linearly independent solutions around the point $z = 0$ are
\begin{equation}
    \phi(z) = \mathcal{A} z^{1-c}{_2F_1}(a+1-c,b+1-c;2-c; z) + \mathcal{B} {_2F_1}(a, b; c; z).
\end{equation}
However, the boundary condition at the outer horizon $(z=0)$ demands an incoming wave solution \cite{cardoso2004small}. Therefore, we can set $\mathcal{B} = 0$ and the resulting radial solution is given by 
\begin{equation}\label{eq:near_sol}
    R(z) = \mathcal{A} z^{-i\hat{\omega}P}(1-z)^{\ell+1}{_2F_1}(a+1-c,b+1-c;2-c;z).
\end{equation}
Since we are interested in the large$-r$ behavior of \eqref{eq:near_sol}, we can use the $z \rightarrow 1-z$ transformation law for the hypergeometric functions \cite{abramowitz1988handbook}:
\begin{equation}
\begin{split}
_2F_1(a-c+1 ,& b-c+1 ; 2-c; z) =
(1-z)^{c-a-b}\frac{\Gamma(2-c)\Gamma(a+b-c)}{\Gamma(a-c+1)\Gamma(b-c+1)} \\
&\times{_2F_1}(1-a,1-b,c-a-b+1,1-z) \\
&+ \frac{\Gamma(2-c)\Gamma(c-a-b)}{\Gamma(1-a)\Gamma(1-b)}{_2F_1}(a-c+1,b-c+1,-c+a+b+1,1-z).
\end{split}
\end{equation}
At large $r$, $z \rightarrow 1$, one has $1-z = \frac{r_+-r_-}{r}$ and $_2F_1(a, b; c; 0)=1$, and thus the asymptotic behavior of the radial solution reads
\begin{equation}\label{eq:near_hor}
R \sim \mathcal{A}\left(\frac{r^\ell}{(r_+-r_-)^\ell}\frac{\Gamma(1-2i\hat{\omega}P)\Gamma(2\ell+1)}{\Gamma(\ell+1)\Gamma(\ell+1-2i\hat{\omega}P)}+\frac{(r_+-r_-)^{\ell+1}}{r^{\ell+1}}\frac{\Gamma(1-2i\hat{\omega}P)\Gamma(-2\ell-1)}{\Gamma(-\ell-2i\hat{\omega}P)\Gamma(-\ell)}\right ).
\end{equation}

\subsubsection{The far-region equation and solution}
\label{sec:3.2.2}
Far from the black hole, the equation of motion simplifies to
\begin{equation}
    \frac{1}{r^2}\frac{d}{dr}\left(r^2\frac{d}{dr}R(r)\right) + \hat{\omega}^2 R(r)-\frac{\ell(\ell+1)}{r^2}R(r)=0,
\end{equation}
where the radial solution can be written in terms of the spherical Bessel functions \cite{sakurai1995modern}: 
\begin{equation}\label{eq:bessel}
    R(r) = \alpha j_{\ell}(\hat{\omega} r) + \beta y_{\ell}(\hat{\omega} r).
\end{equation}
For large $r$, the radial solution \eqref{eq:bessel} becomes
\begin{equation}\label{eq:near_infty}
    R(r) \simeq \alpha \frac{1}{\hat{\omega}r}\cos\left[\hat{\omega}r-\frac{\pi}{2}(\ell+1)\right]+\beta\frac{1}{\hat{\omega}r}\sin\left[\hat{\omega}r-\frac{\pi}{2}(\ell+1)\right].
\end{equation}

\subsubsection{Matching conditions}
\label{sec:3.2.3}

Now, we have to match the radial solutions \eqref{eq:near_sol} and \eqref{eq:bessel} in the overlapping region $M\ll r-r_+ \ll 1/\omega$. In other words, we expand the solution in the far-region for small $r$ and compare it with the near-region solution for large $r$. It turns out that \eqref{eq:near_infty} becomes

\begin{equation} \label{eq:infty_sol}
R(r) \sim \alpha\frac{(\hat{\omega}r)^\ell}{(2\ell+1)!!}-\beta\frac{(2\ell-1)!!}{(\hat{\omega}r)^{\ell+1}}.
\end{equation}

Note that in the low-energy limit $\beta \ll \alpha$, the matching procedure between \eqref{eq:near_hor} and \eqref{eq:infty_sol} gives

\begin{align}\label{eq:coef}
\mathcal{A} &= \frac{\sqrt{\pi}2^{-\ell-1}\hat{\omega}^\ell(r_+-r_-)^\ell\Gamma(\ell+1)\Gamma(\ell+1-2i\hat{\omega}P)}{\Gamma(2\ell+1)\Gamma(\ell+\frac{3}{2})\Gamma(1-2i\hat{\omega}P)}\alpha \nonumber \\
&= \frac{\sqrt{\pi}2^{-\ell-1}\left(\frac{f(E)}{g(E)}\omega\right)^\ell(r_+-r_-)^\ell\Gamma(\ell+1)\Gamma(\ell+1-2i\frac{f(E)}{g(E)}\omega P)}{\Gamma(2\ell+1)\Gamma(\ell+\frac{3}{2})\Gamma(1-2i\frac{f(E)}{g(E)}\omega P)}\alpha.
\end{align}

Now, we can check the change on this relation of the coefficients by assuming a specific form for the rainbow functions, e.g., let us consider \eqref{ex1}. Then, the expression \eqref{eq:coef} yields

\begin{align}
\mathcal{A} &= \frac{\sqrt{\pi}2^{-\ell-1}\left(\frac{e^{E/E_{P}}-1}{E/E_{P}}\omega\right)^\ell(r_+-r_-)^\ell\Gamma(\ell+1)\Gamma(\ell+1-2i\frac{e^{E/E_{P}}-1}{E/E_{P}}\omega P)}{\Gamma(2\ell+1)\Gamma(\ell+\frac{3}{2})\Gamma(1-2i\frac{e^{E/E_{P}}-1}{E/E_{P}}\omega P)}\alpha\\ \nonumber
&= \frac{\sqrt{\pi}\left(\omega\sqrt{M^2-Q^2}\right)^\ell\Gamma(\ell+1)\Gamma(\ell+1-2i\omega P)}{2\Gamma(2\ell+1)\Gamma(\ell+\frac{3}{2})\Gamma(1-2i\omega P)}\alpha \nonumber \\
&\times\biggl[1 +\frac{1}{2}\left(\ell+2i\omega P\psi(1-2i\omega P)-2i\omega P\psi(1+\ell-2i\omega P)\right)\frac{E}{E_{P}}\biggr] +\mathcal{O}\left(\left(\frac{E}{E_{P}}\right)^2\right),
\end{align} 
where $\psi(z)$ is the digamma function defined as $\psi(z)=\frac{d}{dz}\ln\Gamma(z)$. Note that $ r_{\pm} = M\pm\sqrt{M^2-Q^2}$, because $g(E)=1$ in \eqref{ex1}. We can also study the contribution of the rainbow functions \eqref{ex2} as another example. We then obtain
\begin{align}
\mathcal{A} &= \frac{\sqrt{\pi}2^{-\ell-1}\omega^\ell(r_+-r_-)^\ell\Gamma(\ell+1)\Gamma(\ell+1-2i\omega P)}{\Gamma(2\ell+1)\Gamma(\ell+\frac{3}{2})\Gamma(1-2i\omega P)}\alpha \\ \nonumber
&= \frac{\sqrt{\pi}\left(\omega\sqrt{M^2-Q^2}\right)^\ell\Gamma(\ell+1)\Gamma(\ell+1-2i\omega P)}{2\Gamma(2\ell+1)\Gamma(\ell+\frac{3}{2})\Gamma(1-2i\omega P)}\alpha\biggl[1-\frac{\ell Q^{2}}{2(\sqrt{M^2-Q^2})^{2}}\frac{E}{E_{P}}\biggr] + \mathcal{O}\left(\left(\frac{E}{E_{P}}\right)^2\right),
\end{align}
where $\hat{\omega} = \omega$ because $g(E) = f(E)$, but $r_\pm=M\pm\sqrt{M^2-\frac{1}{1-E/E_{P}}Q^2}$. 

In order to calculate the absorption cross section, we introduce the conserved flux associated to the Wronskian of the radial wave equation
\begin{equation}
    J=\frac{2\pi}{i}\left(R^*\Delta\partial_rR-R\Delta\partial_rR^*\right).
\end{equation}
We will find the $s-$wave absorption cross section. By replacing \eqref{eq:near_sol}, the incoming flux at the outer horizon reads
\begin{equation}
    J_{in}=-\frac{\pi\vert\alpha\vert^2}{\hat{\omega}},
\end{equation}
while the absorption flux from infinity given by \eqref{eq:infty_sol} is
\begin{equation}
    J_{abs}=-4\pi\vert\mathcal{A}\vert^2r_+^2\hat{\omega}.
\end{equation}
The absorption cross section is \cite{jung2004proof}
\begin{equation}\label{eq:sigma0}
    \sigma_{abs}^{\ell=0}=\frac{\pi}{f(E)^{2}\omega^2}\frac{J_{abs}^{\ell=0}}{J_{in}^{\ell=0}}=\frac{4\pi r_{+}^{2}}{g(E)^{2}},
\end{equation}
where the additional factor in the denominator is due to gravity's rainbow. Namely, the absorption cross section given by the area of the black hole is corrected by the rainbow function $g(E)$. Let us again analyze \eqref{ex1} and \eqref{ex2} as examples. When we consider \eqref{ex1}, the absorption cross section
\begin{equation}
\sigma_{abs}^{\ell=0}=4\pi\left(M+\sqrt{M^2-Q^2}\right)^2
\end{equation}
reproduces the absorption cross section of the RN black hole without gravity's rainbow because the dependence on $g(E)$ is trivial. On the other hand, when we replace by \eqref{ex2}, it gives 
\begin{equation}\label{re1}
\begin{gathered}
\sigma_{abs}^{\ell=0} = 4\pi r_+^2(1-E/E_{P})^2=4\pi\left(M+\sqrt{M^2-\frac{1}{1-E/E_{P}}Q^2}\right)^2(1-E/E_{P})^2 \\ 
= 4\pi(M+\sqrt{M^2-Q^2})^2-4\pi\biggl(\frac{Q^2(M+\sqrt{M^2-Q^2})}{\sqrt{M^2-Q^2}} \\
+ 2(M+\sqrt{M^2-Q^2})^2\biggr)\frac{E}{E_{P}}+\mathcal{O}\left(\left(\frac{E}{E_{P}}\right)^2\right),
\end{gathered}
\end{equation}
where $E\ll E_{P}$. Note that the first term in \eqref{re1} is the absorption cross section given by the RN black hole, while the effect of gravity's rainbow appears as a linear contribution for $E/E_{P}$. 

\subsection{The Trieste formula}
\label{sec:3.3}

\subsubsection{The Confluent Heun Equation}
\label{sec:3.3.1}
Now we introduce the following M\"{o}bius transformation, and then a s-homotopic transformation
\begin{subequations}
\begin{equation}\label{eq:mobius}
z = \frac{r-r_{-}}{r_{+}-r_{-}},
\end{equation}
\begin{equation}
R(z) = e^{\kappa z}z^{-\theta_{-}/2}(z-1)^{-\theta_{+}/2}\Phi(z),
\end{equation}
\end{subequations}
where the critical exponents are defined by
\begin{equation}
\begin{gathered}
\theta_{-} = i\frac{2\,r_{-}^{2}}{r_{-}-r_{+}}\left(\hat{\omega} - \frac{e Q}{r_{-}}\right), \qquad \theta_{+} = i\frac{2\,r_{+}^{2}}{r_{+}-r_{-}}\left(\hat{\omega} -\frac{e Q}{r_{+}}\right), \\
\kappa = i \hat{\omega} (r_{+}-r_{-}),
\end{gathered}
\end{equation}
to bring the radial equation into the canonical form of the Confluent Heun Equation. The latter reduces \eqref{eq:radial_ODE} to
\begin{equation}\label{eq:confluent_radial}
\Phi^{\prime\prime}(z) + \left[\frac{\gamma}{z}+\frac{\delta}{z-1}+\epsilon\right]\Phi^{\prime}(z) + \frac{\alpha z - q}{z(z-1)}\Phi (z) = 0,
\end{equation}
\begin{equation}
\begin{gathered}
\gamma = 1-\theta_{-}, \qquad \delta = 1-\theta_{+}, \qquad \epsilon = 2\kappa, \\
\alpha =2\kappa\left(1-\theta_{-}-\theta_{+}\right),\\
q = \ell(\ell + 1) + \frac{1}{2}\left(\theta_{-}+\theta_{+}\right) - \kappa\left(2\theta_{-} - 1\right).
\end{gathered}
\end{equation}
The two solutions at $z = 1$ are given by
\begin{equation}
\begin{split}
\Phi(z) &= C_{1}\mathrm{HeunC}(q-\alpha,-\alpha,\delta,\gamma,-\epsilon;1-z) \\
&+ C_{2}(1-z)^{1-\delta}\mathrm{HeunC}(q-\alpha-(1-\delta)(\epsilon+\gamma),-\alpha-(1-\delta)\epsilon,2-\delta,\gamma,-\epsilon;1-z),
\end{split}
\end{equation}
and the asymptotic behavior leads to
\begin{equation}
\Phi(z) \sim \tilde{C}_{1} (z-1)^{-\theta_{+}/2} + \tilde{C}_{2} (z-1)^{\theta_{+}/2},
\end{equation}
which for an incoming wave boundary at the outer horizon $z=1$ requires that $\tilde{C}_{2} = C_{2}\,e^{\kappa} =0$. Therefore, the remaining solution is given by
\begin{equation}\label{eq:incoming_sol}
R(z) = C_{1}e^{\kappa z}z^{-\theta_{-}/2}(z-1)^{-\theta_{+}/2}\mathrm{HeunC}(q-\alpha,-\alpha,\delta,\gamma,-\epsilon;1-z)
\end{equation}
Now, we can express \eqref{eq:incoming_sol} around $z=\infty$ using the formula derived in \cite{Bonelli:2022ten} for the connection coefficients between $z=1,\infty$
\begin{equation}\label{eq:confluent_coeff}
\begin{split}
&\mathrm{HeunC}(q-\alpha,-\alpha,\delta,\gamma,-\epsilon;1-z) \\
&\qquad = \left(\sum_{\sigma=\pm}\frac{\Gamma\left(-2\sigma a(q)\right)\Gamma\left(1-2\sigma a(q)\right)\Gamma\left(\delta\right)\epsilon^{-\frac{1}{2}-\frac{\alpha}{\epsilon}+\frac{\delta+\gamma}{2}+\sigma a(q)}e^{\pm\frac{i\pi}{2}-\frac{1}{2}\partial_{a_{1}}F+\frac{1}{2}\partial_{m}F-\frac{\sigma}{2}\partial_{a}F}}{\Gamma\left(\frac{1-\gamma+\delta}{2}-\sigma a(q)\right)\Gamma\left(\frac{\gamma+\delta-1}{2}-\sigma a(q)\right)\Gamma\left(\frac{1+\gamma+\delta}{2}-\frac{\alpha}{\epsilon}-\sigma a(q)\right)}\right)\\
&\qquad \times z^{-\frac{\alpha}{\epsilon}}\mathrm{HeunC}_{\infty}\left(q,\alpha,\gamma,\delta,\epsilon;z^{-1}\right) \\
&\qquad + \left(\sum_{\sigma=\pm}\frac{\Gamma\left(-2\sigma a(q)\right)\Gamma\left(1-2\sigma a(q)\right)\Gamma\left(\delta\right)\epsilon^{-\frac{1}{2}+\frac{\alpha}{\epsilon}-\frac{\delta+\gamma}{2}+\sigma a(q)}e^{\pm\frac{i\pi}{2}-\frac{1}{2}\partial_{a_{1}}F+\frac{1}{2}\partial_{m}F-\frac{\sigma}{2}\partial_{a}F}}{\Gamma\left(\frac{1-\gamma+\delta}{2}-\sigma a(q)\right)\Gamma\left(\frac{\gamma+\delta-1}{2}-\sigma a(q)\right)\Gamma\left(\frac{1-\gamma-\delta}{2}+\frac{\alpha}{\epsilon}-\sigma a(q)\right)}\right)\\
&\qquad \times e^{-\epsilon z}z^{\frac{\alpha}{\epsilon}-\gamma-\delta}\mathrm{HeunC}_{\infty}\left(q-\gamma\epsilon,\alpha-\epsilon(\gamma+\delta),\gamma,\delta,-\epsilon;z^{-1}\right),
\end{split}
\end{equation}
where $a(q)$ is obtained by inverting the Matone relation. $F$ is the classical confluent conformal block which via the AGT correspondence can be expressed in terms of the instanton NS free energy $\mathcal{F}^{\mathrm{inst}}$, defined in \eqref{eq:Finst}.
By replacing \eqref{eq:mobius} and then, expanding the solution it around $r = \infty$, the asymptotic behavior yields
\begin{equation}
\begin{split}
R(r) \sim\, &C_{1} \Xi_{1} \left(r_{+}-r_{-}\right)^{1-i\left(\hat{\omega}(r_{+}+r_{-})-e\,Q\right)}e^{i\hat{\omega} r}r^{-1+i\left(\hat{\omega}(r_{+}+r_{-}) - e\,Q\right)} \\
&\qquad + C_{1} \Xi_{2} \left(r_{+}-r_{-}\right)^{1+i\left(\hat{\omega}(r_{+}+r_{-})-e\,Q\right)}e^{-i\hat{\omega} r}r^{-1-i\left(\hat{\omega}(r_{+}+r_{-})-e\,Q\right)},
\end{split}
\end{equation}
where $\Xi_{1,2}$ are the $r-$independent terms of the connection coefficients in \eqref{eq:confluent_coeff}. By inspecting the leading behavior of the asymptotic solutions, we notice that at infinity the radial coefficients in \eqref{eq:boundaryforr} are defined as
\begin{equation}
A_{in} =  C_{1} \Xi_{2} \left(r_{+}-r_{-}\right)^{1+i\left(\hat{\omega}(r_{+}+r_{-})-e\,Q\right)}, \qquad A_{out} = C_{1} \Xi_{1} \left(r_{+}-r_{-}\right)^{1-i\left(\hat{\omega}(r_{+}+r_{-}) - e\,Q\right)},
\end{equation}
where the horizons $r_{\pm}$ are given by \eqref{eq:horizons}.

\subsubsection{The radial dictionary}
\label{sec:3.3.2}

Consider the following transformation

\begin{equation}\label{eq:SL_transorfmation}
R(z) = z^{-1/2}(z-1)^{-1/2}\psi_{\omega,\ell}(z),
\end{equation}
which takes the radial equation \eqref{eq:radial_ODE} to the $\mathrm{SL}(2,\mathds{C})$ form
\begin{equation}
\frac{d^{2}\psi(z)}{dz^{2}} + \left(\frac{\frac{1}{4}-a_{0}^{2}}{z^{2}} + \frac{\frac{1}{4}-a_{1}^{2}}{(z-1)^{2}} + \frac{u - \frac{1}{2}+a_{0}^{2}+a_{1}^{2}}{z(z-1)} - \frac{m_{3} \Lambda}{z} - \frac{\Lambda^{2}}{4}\right)\psi(z) = 0,
\end{equation}
and by comparing with \eqref{eq:confluent_radial} we can establish the following dictionary
\begin{equation}\label{eq:dictionary}
\begin{split}
&a_{0} = \pm i \frac{r_{-}^{2}}{r_{-}-r_{+}}\left(\hat{\omega} - \frac{e\,Q}{r_{-}}\right), \\
&a_{1} = \pm i \frac{r_{+}^{2}}{r_{+}-r_{-}}\left(\hat{\omega} - \frac{e\,Q}{r_{+}}\right), \\
&m_{3} = \pm i \left((r_{+} + r_{-}) \hat{\omega} - e\,Q\right), \\
&\Lambda = \pm i 2 \hat{\omega} (r_{+}-r_{-}), \\
&u = \left(\hat{\omega}(r_{+} + r_{-}) - e\,Q\right)^{2} + 2\hat{\omega} r_{+}^{2}\left(\hat{\omega} - \frac{e\,Q}{r_{+}}\right) - \ell(\ell+1),
\end{split}
\end{equation}
depending on the choice of the sign in \eqref{eq:dictionary}. Furthermore, $u$ can be defined via the Matone relation
\begin{equation}\label{eq:Matone}
u = \frac{1}{4} - a^{2} + \Lambda \partial_{\Lambda}\mathcal{F}^{\rm inst},
\end{equation}
where $\mathcal{F}^{\rm inst}$ is the instanton part of the NS free energy given in \eqref{eq:Finst}. The masses in the dual gauge theory are given by
\begin{equation}\label{eq:masses}
\begin{split}
&m_{1} = a_{0} + a_{1},\\  
&m_{2} = a_{1} - a_{0},\\
&m_{3} = i \left(\omega(r_{+} + r_{-}) - e\,Q\right),
\end{split}
\end{equation}
which are purely imaginary and correspond to physical Liouville momenta.

\subsubsection{Absorption cross section in the low-energy limit}
\label{sec:3.3.3}

The analytic expansions in \eqref{eq:derivatives} require that $\hat{\omega} (r_{+}-r_{-}) < 1$ to ensure their convergence, which corresponds to a low-energy limit. Bearing this in mind we define
\begin{equation}
\epsilon \equiv r_{+} - r_{-},
\end{equation}
so that by inverting \eqref{eq:Matone} and replacing $a(q)$ by the following Taylor series
\begin{equation}
a = \sum_{n}a_{n}\hat{\omega}^{n},
\end{equation}
we can compute the coefficients recursively. The asymptotic expansion for the Floquet exponent $a$ is of the form
\begin{equation}\label{eq:sigma}
a =  \ell + \frac{1}{2} + \left(\frac{6 \epsilon r_{+}}{1 + 2\ell} - \frac{6 r_{+}^{2}}{1 + 2\ell} - \frac{(15\ell^{2}-15\ell-11)\epsilon^{2}}{2(1 + 2\ell)(2\ell - 1)(3 + 2\ell)}\ \right)\hat{\omega}^{2} + \mathcal{O}(\hat{\omega}^{4}),
\end{equation}
where we have assumed $e = 0$. Recently, a similar expansion has been computed in \cite{Bianchi:2023sfs} using an infite fraction equation. Following the analysis in \cite{maldacena1997universal,Bonelli:2021uvf}, one can define the absorption cross section as the ratio of the conserved fluxes associated to the radial equation \eqref{eq:radial_ODE}, which can be written in terms of the connection coefficients of the confluent Heun functions \eqref{eq:confluent_coeff}
\begin{equation}\label{eq:cross_section}
\sigma^{\ell} = \frac{\pi}{f(E)^{2}\omega^{2}}\frac{J^{\ell}_{abs}}{J^{\ell}_{in}} = \frac{\pi}{f(E)^{2}\omega^{2}}\frac{r_{+}^{2}}{\epsilon^{2}}\frac{1}{\vert \Xi_{2}\vert^{2}},
\end{equation}
where $J^{\ell}_{abs}$ and $J^{\ell}_{in}$ are the flux across the horizon and the incoming flux from infinity, respectively. In order to compute \eqref{eq:cross_section}, we consider terms up to fourth order for small $\hat{\omega}$, while fixing $\epsilon < r_{+}$. Then, by expanding \eqref{eq:dictionary} at the low-energy limit
\begin{equation}\label{eq:small}
\begin{split}
&a_{0} = -i\frac{\left(r_{+}-\epsilon\right)^{2}}{\epsilon}\hat{\omega}, \\
&a_{1} = \frac{i \hat{\omega} r_{+}^{2}}{\epsilon}, \\
&m_{3} = i\left(2r_{+} - \epsilon\right)\hat{\omega}, \\
&\Lambda = 2i\hat{\omega}\epsilon, \\
&\alpha = 2i\hat{\omega}\epsilon + 4\epsilon\left(2r_{+}-\epsilon\right)\hat{\omega}^{2},
\end{split}
\end{equation}
and taking into account \eqref{eq:sigma}, the absorption cross section \eqref{eq:cross_section} for the $\ell = 0$ case reads
\begin{equation}\label{eq:seccion}
\begin{split}
\sigma^{0}_{abs} &= \frac{4\pi r_{+}^{2}}{g(E)^{2}}\biggl\lbrace 1 - \pi \hat{\omega} \left(\frac{2r_{+}^{2}}{\epsilon} + 2 r_{+} - \epsilon\right)  + \biggl[\frac{2\pi^{2}r_{+}^{2}}{\epsilon^{2}}\left(r_{+}^{2}+\frac{8}{3}r_{+}\epsilon - \frac{4}{3}\epsilon^{2}\right) \\
&+ \left(26r_{+}^{2} - 22r_{+}\epsilon +\frac{227}{36}\epsilon^{2}\right) -\frac{1}{3}\left(36r_{+}^{2}-36r_{+}\epsilon + 11\epsilon^{2}\right)\left(\gamma + \log 2\hat{\omega}\epsilon \right)\biggr]\hat{\omega}^{2}\biggr\rbrace + \mathcal{O}(\hat{\omega}^{3}),
\end{split}
\end{equation}
where $\gamma$ is the Euler-Mascheroni constant. The expression \eqref{eq:seccion} reduces to \eqref{eq:sigma0} for $\hat{\omega} \rightarrow 0$, giving the area of the RN black hole. It is worth mentioning that for $\epsilon \rightarrow 0$, although $a_{0}, a_{1} \rightarrow \infty$, the Nekrasov-Shatashvili free energy will develop a confluent limit due to $\Lambda \rightarrow 0$, which compensates the latter divergences, as it has been pointed out in \cite{Cavalcante:2021scq}. 

\section{Conclusions}
\label{sec:4}
We have investigated the low-energy absorption cross section by RN black hole in gravity's rainbow. More precisely, we have followed two approaches to compute the absorption cross section. In the overlapping method, we divide the physical problem into two regions: The first region is near the horizon, while the second is far from the black hole. In both cases, the radial equation simplifies and can be solved in terms of the hypergeometric functions and spherical Bessel functions, respectively. To match these solutions, we expand the solution in the far-region \eqref{eq:infty_sol} for small $r$ and compare it with the near-region solution \eqref{eq:near_sol} for large $r$. In the low-energy limit, the ratio between the flux across the horizon and the flux incoming from infinity reproduces the area of the horizon. The second approach explores the solution of the radial confluent Heun equation \eqref{eq:confluent_radial} and its connection coefficients \eqref{eq:confluent_coeff} to derive an expression for the absorption cross section \eqref{eq:cross_section}. We obtained an asymptotic expansion for the absorption cross section in the low-energy limit \eqref{eq:sigma0}. The overlapping method can be thought of as the zeroth order contribution compared with the result coming from the connection coefficients of the full radial ODE.

In gravity's rainbow, the frequency is transformed in principle as $\omega\rightarrow f(E)\omega$ because it is associated with the $0-$component of the four-momentum of the scalar field. However, the transformed radial coordinate contains the rainbow function $g(E)$ in the denominator, modifying the frequency as 
\begin{equation}\label{eq:freq}
\omega\rightarrow\frac{f(E)}{g(E)}\omega,
\end{equation}
in the radial equation. As a result, the absorption cross section for $\ell=0$ is the horizon area divided by the rainbow function \eqref{eq:sigma0}. Nevertheless, this result reproduces the $s$-wave absorption cross section in GR for $g(E) = 1$. We have observed the effects of gravity's rainbow by assuming specific forms for the rainbow functions, which are of linear order in $E/E_{P}$. Exploring the QNMs spectrum under gravity's rainbow seems like a natural step and will be addressed in future work.

\begin{acknowledgments}
We would like to thank Alba Grassi, Bruno Carneiro da Cunha, Giulio Bonelli and Francisco Brito for stimulating and helpful discussions. This research was supported by Basic Science Research Program through the National Research Foundation of Korea (NRF) funded by the Ministry of Education (NRF-2022R1I1A2063176) and the Dongguk University Research Fund of 2023. BG appreciates APCTP for its hospitality during the topical research program, {\it Multi-Messenger Astrophysics and Gravitation}.
\end{acknowledgments}

\appendix
\section{Nekrasov partition function}
\label{appendix_A}
The irregular conformal blocks can be computed via the AGT correspondence as a gauge theory instanton partition function \cite{Alday:2009aq},
\begin{equation}
\braket{\Delta_{\alpha},\Lambda_{0},m_{0} \vert V_{\alpha_{1}}(1) \vert \Delta_{\alpha_{0}}} = \mathcal{Z}_{SU(2)}^{\rm inst}(\Lambda,a,m_{1},m_{2},m_{3}),
\end{equation}
where $\mathcal{Z}^{\rm inst}$ is the Nekrasov instanton partition function of $\mathcal{N}=2$ $SU(2)$ gauge theory with three hypermultiplets. The $SU(2)$ partition function is given by the $U(2)$ partition function divided by the $U(1)$-factor,
\begin{equation}
\mathcal{Z}_{SU(2)}^{\rm inst}(\Lambda,a,m_{1},m_{2},m_{3},\epsilon_{1},\epsilon_{2}) = \mathcal{Z}_{U(1)}^{-1}(\Lambda,m_{1},m_{2},\epsilon_{1},\epsilon_{2})\mathcal{Z}_{U(2)}^{\rm inst}(\Lambda,a,m_{1},m_{2},m_{3},\epsilon_{1},\epsilon_{2}),
\end{equation}
where the $U(2)$ partition function is expressed in terms of the following combinatorial formula. Let $Y = \left(\lambda_{1} \geq \lambda_{2} \geq \ldots \right)$ be a Young tableau where $\lambda_{i}$ is the height of the i-th column and we set $\lambda_{i} = 0$ when $i$ is larger than the width of the tableau. Its transpose is denoted by $Y^{T} = \left(\lambda^{\prime}_{1} \geq \lambda^{\prime}_{2} \geq \ldots \right)$. For a box $s$ at the coordinate $(i,j)$ we define the arm-length $A_{Y}(s)$ and the leg-length $L_{Y}(s)$ with respect to the tableau $Y$ as
\begin{equation}
A_{Y}(s) = \lambda_{i} - j, \qquad L_{Y}(s) = \lambda^{\prime}_{j} - i.
\end{equation}
Define the function $E$ by
\begin{equation}
E(a,Y_{1},Y_{2},s) = a - \epsilon_{1}L_{Y_{2}}(s) + \epsilon_{2}\left(A_{Y_{1}}(s) + 1 \right).
\end{equation}
Using the notation $\vec{a} = \left(a_{1},a_{2}\right)$ with $a_{1} = - a_{2} = a$ and $\vec{Y} = \left(Y_{1},Y_{2}\right)$, the contribution of a vector multiplet reads
\begin{equation}
z^{\rm inst}_{\rm vector}(\vec{a},\vec{Y}) = \prod_{i,j=1}^{2}\prod_{s \in Y_{i}}\frac{1}{E(a_{i}-a_{j},Y_{i},Y_{j},s)}\prod_{t \in Y_{j}}\frac{1}{\epsilon_{1} + \epsilon_{2} -E(a_{j}-a_{i},Y_{j},Y_{i},t)}
\end{equation}
and that of a matter hypermultiplet
\begin{equation}
z^{\rm inst}_{\rm matter}(\vec{a},\vec{Y},m) = \prod_{i=1}^{2}\prod_{s \in Y_{i}}\left(a + m +\epsilon_{1}\left(i - \frac{1}{2}\right) + \epsilon_{2}\left(j - \frac{1}{2}\right)\right).
\end{equation}
Finally, the $U(2)$ partition function is given by
\begin{equation}
\mathcal{Z}^{\rm inst}_{U(2)}(\Lambda,a,m_{1},m_{2},m_{3}) = \sum_{\vec{Y}}\Lambda^{\vert \vec{Y} \vert} z^{\rm inst}_{\rm vector}(\vec{a},\vec{Y}) \prod_{n=1}^{3} z^{\rm inst}_{\rm matter}(\vec{a},\vec{Y},m_{n}),
\end{equation}
where $\vert \vec{Y} \vert$ denotes the total number of boxes in $Y_{1}$ and $Y_{2}$. The $U(1)$ partition function with $N_{f}=3$ can be obtained by decoupling one mass from the $U(1)$-factor for $N_{f}=4$, or in other words, by the coalescence of two vertex operators to form the irregular state. By taking such limit, we get
\begin{equation}
\mathcal{Z}_{U(1)} =  e^{-\left(m_{1} + m_{2} + \epsilon\right)\Lambda/2\epsilon_{1}\epsilon_{2}},
\end{equation} 
where $\epsilon = \epsilon_{1} + \epsilon_{2}$, $\epsilon_{i}$ are two $\Omega$-background parameters regulating the infrared divergence in the localization computation. In this paper we work in the Nekrasov-Shatashvili limit which is defined by $\epsilon_{2} \rightarrow 0$ and $\epsilon_{1}=1$. The instanton part of the NS free energy is defined as

\begin{equation}\label{eq:Finst}
\mathcal{F}^{\rm inst}(\Lambda,a,m_{1},m_{2},m_{3},\epsilon_{1}) = \epsilon_{1}\lim_{\epsilon_{2} \rightarrow 0}\log \mathcal{Z}^{\rm inst}_{SU(2)}(\Lambda,a,m_{1},m_{2},m_{3},\epsilon_{1}).
\end{equation}
For reference, we compute the relevant quantities for the connection coefficients

\begin{equation}\label{eq:derivatives}
\begin{split}
&\partial_{a_{1}}\mathcal{F}^{\rm inst} = \frac{(m_{1}+m_{2})}{\frac{1}{2}-2a^{2}}m_{3}\Lambda + \mathcal{O}(\Lambda^{2}) \\
&\partial_{a}\mathcal{F}^{\rm inst} = \left(\frac{a}{\frac{1}{4} - a^{2}} - \frac{a\left(\frac{1}{4} - a^{2} - m_{1}m_{2}\right)}{\left(\frac{1}{4} - a^{2}\right)^{2}}\right) m_{3}\Lambda + \mathcal{O}(\Lambda^{2}) \\
&\partial_{m_{3}}\mathcal{F}^{\rm inst} = -\frac{\left(\frac{1}{4} - a^{2} - m_{1}m_{2}\right)}{\frac{1}{2} - 2a^{2}}\Lambda + \mathcal{O}(\Lambda^{2})
\end{split}
\end{equation}

\bibliographystyle{unsrt}
\bibliography{ref_raingrav}
\end{document}